\documentclass[aps,amssymb,amsmath,pra,superscriptaddress]{revtex4}
\usepackage{graphicx}
\usepackage{epsfig}
\usepackage{dcolumn}% Align table columns on decimal point
\usepackage{bm}% bold math
\usepackage{bbm}
\usepackage{amsmath}
\usepackage{graphicx}
\usepackage{amsfonts}
\usepackage{hyperref}
\usepackage{amssymb}
\usepackage{color}
\usepackage[up]{subfigure}

\newcommand{\be}{\begin{equation}}
\newcommand{\ee}{\end{equation}}
\newcommand{\bc}{\begin{center}}
\newcommand{\ec}{\end{center}}
\newcommand{\bea}{\begin{eqnarray}}
\newcommand{\eea}{\end{eqnarray}}

\begin{document}
\title{Quantumness     in     decoherent     quantum    walk     using
  measurement-induced disturbance}

\author{R. \surname{Srikanth}}
\email{srik@poornaprajna.org}
\affiliation{Poornaprajna Institute of Scientific Research,
Devanahalli, Bangalore 562 110, India}
\affiliation{Raman Research Institute, Sadashiva Nagar, Bangalore, India}
\author{Subhashish \surname{Banerjee}}
\email{subhashish@cmi.ac.in}
\affiliation{Chennai Mathematical Institute, Siruseri, Chennai, India}
\author{C. M. \surname{Chandrashekar}}
\email{cmadaiah@iqc.ca ; chandru@imsc.res.in}
\affiliation{Institute for Quantum Computing, University of Waterloo, Waterloo, Ontario, Canada N2L 3G1}
\affiliation{Center for Quantum Sciences, The Institute of Mathematical Sciences, Chennai 600113, India}

%==================================================

\begin{abstract}
The classicalization  of a decoherent discrete-time quantum  walk on a
line or an  $n$-cycle can be demonstrated in various  ways that do not
necessarily provide a  geometry-independent description.  For example,
the position  probability distribution becomes  increasingly Gaussian,
with a concomitant fall in the standard deviation, in the former case,
but not in  the latter.  As another example, each  step of the quantum
walk on  a line may be  subjected to an arbitrary  phase gate, without
affecting the position probability distribution, no matter whether the
walk is noiseless  or noisy.  This {\em symmetry},  which is absent in
the case of noiseless cyclic walk,  but is restored in the presence of
sufficient noise, serves as an indicator of classicalization, but only
in  the  cyclic  case.   Here  we  show that  the  degree  of  quantum
correlations  between  the  coin  and  position  degrees  of  freedom,
quantified  by a  measure based  on the  disturbance induced  by local
measurements [Luo, Phys.  Rev.  A {\bf 77}, 022301 (2008)], provides a
suitable   measure   of   classicalization   across   both   type   of
walks. Applying  this measure to compare  the two walks,  we find that
cyclic quantum walks tend to classicalize faster than quantum walks on
a  line because  of  more  efficient phase  randomization  due to  the
self-interference of  the two counter-rotating waves.   We model noise
as acting on the coin, and given by the squeezed generalized amplitude
damping  (SGAD) channel, which  generalizes the  generalized amplitude
damping channel.
\end{abstract}

%====================================================

\maketitle

%================================================
\section{Introduction}
\label{sec:intro}
%================================================

Quantum walks  (QWs) are  the quantum analog  of the  classical random
walks (CRWs)  \cite{Ria58, FH65, ADZ93,  DM96, FG98}.  It is  now well
known that the quantum features of interference and superposition make
QW  spread  quadratically  faster  than  CRW.   Like  their  classical
counterpart,  QWs  are  also   widely  studied  in   two  forms:
continuous-time  QW  (CTQW) \cite{FG98}  and  discrete-time QW  (DTQW)
\cite{ADZ93, DM96, ABN01, NV01} and are found to be very useful from the perspective
of quantum algorithms \cite{Amb03, CCD03, SKB03, AKR05}.  Furthermore,
they have been  used to demonstrate the coherent  quantum control over
atoms, quantum phase transition \cite{CL08}, to explain phenomena such
as the  breakdown of an electric-field driven  system \cite{OKA05} and
direct  experimental  evidence  for  wavelike energy  transfer  within
photosynthetic    systems     \cite{ECR07,    MRL08}.     Experimental
implementation of  QWs has  been reported with  samples in  an nuclear
magnetic resonance (NMR) system \cite{DLX03, RLB05}, in the continuous
tunneling of light fields  through waveguide lattices \cite{PLP08}, in
the  phase space  of trapped  ions  \cite{SMS09, ZKG10}  based on  the
scheme proposed by \cite{TM02}, with single neutral atoms in optically
trapped  atoms  \cite{KFC09}, and  with  single photons  \cite{BFL10}.
Recently, implementation  of a QW-based  search algorithm in  an NMR system
has  been  reported \cite{LZZ10}.   Various  other  schemes have  been
proposed for  their physical  realization in different  physical systems
\cite{RKB02, EMB05, Cha06, MBD06}.

Decoherence in QW  and the transition of QW to  CRW is quite important
from the  viewpoint of practical implementation, and  has been studied
by various authors \cite{KT03,  BCA03, Ken06, CSB07, BSC08, LP10}.  In
particular,  in  Refs.   \cite{CSB07,  BSC08},  we  investigated  some
qualitatively different  ways in which  environmental effects suppress
quantum superposition  in QW on  a line and  an $n-$ cycle.   In these
works, we have modeled noise as  the bit flip channel, the phase flip
channel  (decoherence   without  dissipation),  and   the  generalized
amplitude damping channel (decoherence with dissipation), which are of
relevance to  studies in quantum optics and  condensed matter systems.
The latter type of noise in  the QW context appears to have been first
studied by us \cite{CSB07}. Here we employ a more general noise acting
on  the  coin,  the  squeezed  generalized  amplitude  damping  (SGAD)
channel, that  arises due  to coupling with  a squeezed,  thermal bath
\cite{SB08}.

In  this work,  we consider  the issue  of quantifying  the  degree of
quantumness  of  the walker  during  the  transition  from a  pure  QW
dynamics  to the CRW  dynamics.  As  a general  problem applied  to an
arbitrary  quantum  system,  this   is  still  not  fully  understood.
Intuitively, one  would expect that quantum  entanglement captures the
entire content of  quantum correlations in a system,  but this may not
be so, in  view of other contributions, coming  from superposition and
noncommutativity  of observables.  In Ref.   \cite{OZ01},  a quantity
called  `discord', which  is  defined as  the  difference between  two
different quantum generalizations  of classical mutual information, is
shown to be  a viable measure of the quantumness  of correlations in a
bipartite  system.   One  can  produce  systems in  which  discord  is
nonvanishing even  though entanglement vanishes.   However, computing
discord  in a  large-dimensional system  can be  difficult  because it
requires   an  optimization  over   all  possible   local  measurement
strategies. An alternative method  is via `dissonance', defined as
  the  relative entropy  between a  separable state  $\sigma$  and the
  closest {\em classical  state} (a separable state that  is a mixture
  of locally  distinguishable states), where $\sigma$ is  such that it
  has the smallest relative entropy of entanglement with the given (in
  general, entangled)  state $\rho$ \cite{MPSVW10}.  Therefore, unlike
  discord,  dissonance  is  a  measure of  quantumness  that  excludes
  entanglement. An operational approach    that    employs
measurement-induced disturbance  as a measure of  quantumness has been
proposed by  Luo \cite{L08}.  In  this work, we use  this `quantumness
via  measurement-induced  disturbance'  (QMID)  between the  coin  and
position degrees of freedom to study the effect of decoherence on QW.

\par  This article  is arranged  as  follows. In  Sec.  \ref{dtqw}  we
describe  the  DTQW  model  on  a  line and  an  $n-$cycle.   In  Sec.
\ref{sec:envb}  we introduce  the  SGAD channel  as  the process  that
decoheres   the   walk,   and   discuss   some   signatures   of   the
classicalization of QW. In Sec.  \ref{sec:symm}, we discuss a specific
kind  of  walk  symmetry  as  a  means to  characterize  the  loss  of
quantumness in  a decoherent  walk, and point  out its  drawbacks.  In
Sec.  \ref{MID} we investigate QMID as a means to quantify quantumness
of  the correlations  between the  coin  and the  position degrees  of
freedom, that is applicable to decoherent QWs both on a line and on an
$n-$cycle.  Finally, we conclude in Sec. \ref{conclusion}.

%===================================
\section{Discrete-time quantum walk on a line and an $n-$cycle}
\label{dtqw}
%========================================

The DTQW  in  one dimension is  modeled  as  a  particle consisting  of  a
two-level coin  (a qubit)  living in the  Hilbert space  ${\cal H}_c$,
spanned  by $|0\rangle$  and  $|1\rangle$, and  a  position degree  of
freedom  living  in  the   Hilbert  space  ${\cal  H}_p$,  spanned  by
$|\psi_x\rangle$, where $x \in {\mathbbm  I}$, the set of integers. In
an  $n$-cycle walk,  there  are  only $n$  allowed  positions, and  in
addition  the   periodic  boundary  condition  $|\psi_x\rangle=|\psi_{x
\;{\rm mod}\;n}\rangle$ is imposed.  A  $t$-step coined QW is generated
by  iteratively applying  a unitary  operation $W$  which acts  on the
Hilbert     space    ${\cal     H}_c\otimes    {\cal     H}_p$:    \be
|\Psi_t\rangle=W^t|\Psi_{in}\rangle,  \ee  where $|\Psi_{in}\rangle  =
(\cos(\delta        /2)|0\rangle+       \sin(\delta/2)       e^{i\phi}
|1\rangle)\otimes |\psi_0\rangle$  is  an   arbitrary  initial  state  of  the
particle and $W\equiv U(B \otimes  {\mathbbm 1})$, where $U(2) \ni B =
B_{\xi,\theta,\zeta}       \equiv      \left(      \begin{array}{clcr}
  \mbox{~}e^{i\xi}\cos(\theta)      &     &     e^{i\zeta}\sin(\theta)
  \\ e^{-i\zeta} \sin(\theta) & & - e^{-i\xi}\cos(\theta)
\end{array} \right)$ (with components
denoted  $B_{jk}$) is  the  coin operation.   $U$ is  the controlled-shift
operation            \be            U\equiv           |0\rangle\langle
0|\otimes\sum_x|\psi_x-1\rangle\langle   \psi_x|   +  |1\rangle\langle
1|\otimes\sum_x|\psi_x+1\rangle\langle \psi_x|.   \ee For an $n-$cycle
$U$  is  replaced by  \be  U^{c}  \equiv  |0\rangle \langle  0|\otimes
\sum_{x=0}^{n-1}|\psi_{x-1~{\rm  mod}~n}\rangle   \langle  \psi_{x}  |
+|1\rangle   \langle  1   |\otimes   \sum_{x=0}^{n-1}  |\psi_{x+1~{\rm
    mod}~n}\rangle \langle  \psi_{x}| \ee The probability  to find the
particle at  site $x$ after  $t$ steps is  given by $p(x,t)  = \langle
\psi_x|{\rm tr}_c (|\Psi_t\rangle\langle\Psi_t|)|\psi_x\rangle$.

%========================================
\section{Decoherence via squeezed generalized
amplitude damping channel \label{sec:envb}}
%========================================

A  QW implemented  on a  quantum  computer is  inevitably affected  by
errors  caused  by noise  due  to  the  environment.  Some  physically
relevant  models  of  noise  are as follows:  a  phase  flip  channel  (which  is
equivalent to a phase damping or purely dephasing channel), a bit flip
channel, a  generalized amplitude damping  channel ($T \ge 0$)  and an
SGAD  channel, which is  used here.   Our numerical  implementation of
these channels evolves the density  matrix at each walk step employing
the Kraus operator representation for them.

The physical origin and the effect  of various types of noise, such as
phase flip, bit flip,  amplitude and generalized amplitude damping, on
the DTQW  have been  studied in detail  in Refs.   \cite{CSB07, BSC08}.
Here we study  the behavior of decoherent QW  subjected to a recently
introduced  noise  process, the  SGAD  channel  \cite{SB08}, which  we
briefly discuss below for completeness.

System evolution in the interaction picture, for the case of SGAD, has
the following form \cite{SZ97,BP02}:
\begin{eqnarray}
{d \over dt}\rho^s(t) &=& \gamma_0 (N + 1) \left(\sigma_-  \rho^s(t)
\sigma_+ - {1 \over 2}\sigma_+ \sigma_- \rho^s(t) -
{1 \over 2} \rho^s(t) \sigma_+ \sigma_- \right) \nonumber\\
& + & \gamma_0 N \left( \sigma_+  \rho^s(t)
\sigma_- - {1 \over 2}\sigma_- \sigma_+ \rho^s(t) -
{1 \over 2} \rho^s(t) \sigma_- \sigma_+ \right) \nonumber\\
& - & \gamma_0 M   \sigma_+  \rho^s(t) \sigma_+ -
\gamma_0 M^* \sigma_-  \rho^s(t) \sigma_-. 
\label{4b} 
\end{eqnarray}

Here  $\gamma_0  = (4  \omega^3  |\vec{d}|^2)/(3  \hbar  c^3)$ is  the
spontaneous  emission  rate,  while  $\sigma_+$,  $\sigma_-$  are  the
standard  raising   and  lowering   operators.   Also,  $N   =  N_{\rm
  th}[\cosh^2(r)  +  \sinh^2(r)]  +  \sinh^2(r)$,  $M  =  -\frac{1}{2}
\sinh(2r)  e^{i\Phi}  (2  N_{\rm  th}  +  1)$,  where  $N_{\rm  th}  =
1/(e^{\hbar \omega_0/k_B T} - 1)$ is the Planck distribution giving the
number of  thermal photons at frequency $\omega_0$,  while $r$, $\Phi$
are  the   bath  squeezing  parameters  \cite{CS85}.

The SGAD channel is generated  by the master Eq. (\ref{4b}), with
nonvanishing  $r, \Phi,  T$.  The Kraus  operators for  this channel,
depicting  the system-bath  interaction for  a time  $t =  \Delta$ are
\cite{SB08}
\begin{eqnarray}
\begin{array}{ll}
E_0 \equiv \sqrt{p_1}\left[\begin{array}{ll} 
\sqrt{1-\alpha( \Delta)} & 0 \\ 0 & 1
\end{array}\right]; ~~~~ &
E_1 \equiv \sqrt{p_1}\left[\begin{array}{ll} 0 & 0 
\\ \sqrt{\alpha( \Delta)} & 0 \end{array}\right];  \\
E_2 \equiv \sqrt{p_2}\left[\begin{array}{ll} 
\sqrt{1-\mu( \Delta)} & 0 \\ 0 & \sqrt{1-\nu( \Delta)}
\end{array}\right]; ~~~~ &
E_3 \equiv \sqrt{p_2}\left[\begin{array}{ll} 0 & \sqrt{\nu( \Delta)} 
\\ \sqrt{\mu(\Delta)}e^{-i\theta( \Delta)}  & 0
\end{array}\right].\nonumber  
\end{array}
\label{eq:new0}
\end{eqnarray}

Here,
\begin{eqnarray}
\nu( \Delta) &=& \frac{N}{p_2(2N+1)}(1-e^{-\gamma_0(2N+1) \Delta}), \nonumber \\
\mu( \Delta) &=& \frac{2N+1}{2p_2 N}
\frac{\sinh^2(\gamma_0a \Delta/2)}{\sinh(\gamma_0(2N+1) \Delta/2)}
\exp\left(-\frac{\gamma_0}{2}(2N+1) \Delta\right), \nonumber \\
\alpha(\Delta) &=& \frac{1}{p_1}\left(1 - p_2[\mu(\Delta)+\nu(\Delta)]
- e^{-\gamma_0(2N+1)\Delta}\right), \nonumber
\end{eqnarray}
$$ p_1 + p_2 = 1, $$
\begin{eqnarray}
p_2(\Delta) &=& \frac{1}{(A+B-C-1)^2-4D}
 \times \{A^2B + C^2 + A[B^2 - C - B(1+C)-D] - (1+B)D - C(B+D-1) 
\nonumber  \\
&\pm&  2\sqrt{D[B-AB+(A-1)C+D][A-AB+(B-1)C+D]} \},
\label{eq:p2} \nonumber
\end{eqnarray}
where
$A = \frac{2N+1}{2N} \frac{\sinh^2(\gamma_0 a\Delta/2)}
{\sinh(\gamma_0(2N+1)\Delta/2)}
\exp\left(-\gamma_0(2N+1)\Delta/2\right), 
B = \frac{N}{2N+1}(1-\exp(-\gamma_0(2N+1)\Delta)), 
C =  A + B + \exp(-\gamma_0 (2N+1)\Delta), 
D = \cosh^2(\gamma_0 a\Delta/2)\exp(-\gamma_0(2N+1)\Delta).$

The  SGAD  channel extends  the  concept  of  a generalized  amplitude
damping (GAD) channel, which arises due to the dissipative interaction
with a purely thermal bath to  that with a squeezed thermal bath.  One
recovers  generalized  amplitude  damping  channel, if  the  squeezing
parameter $r \rightarrow 0$.

\begin{figure}
\subfigure[]{\label{fig:sqw45} \includegraphics[width=8.8cm]{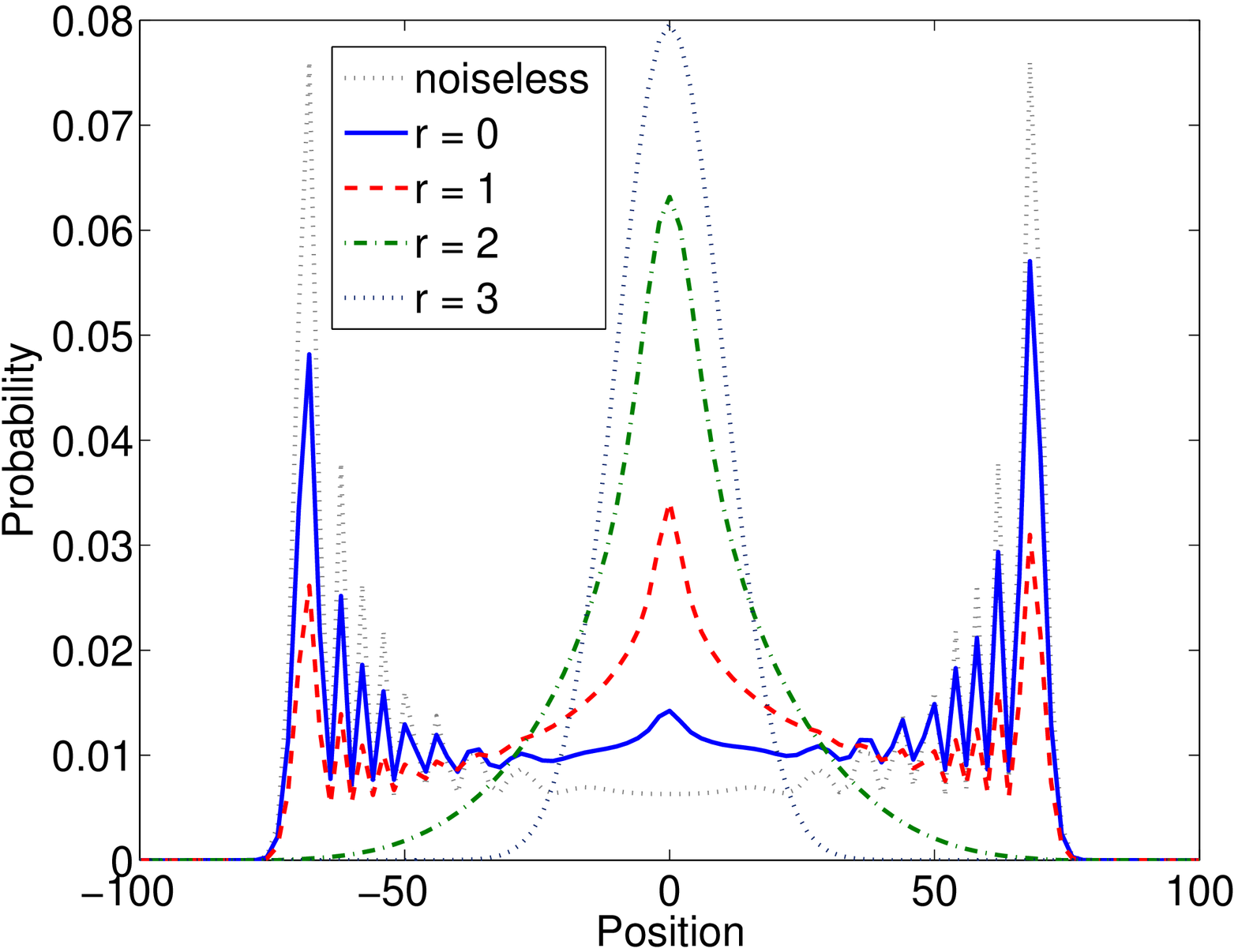}}
\hfill
\subfigure[]{\label{fig:sd} \includegraphics[width=8.8cm]{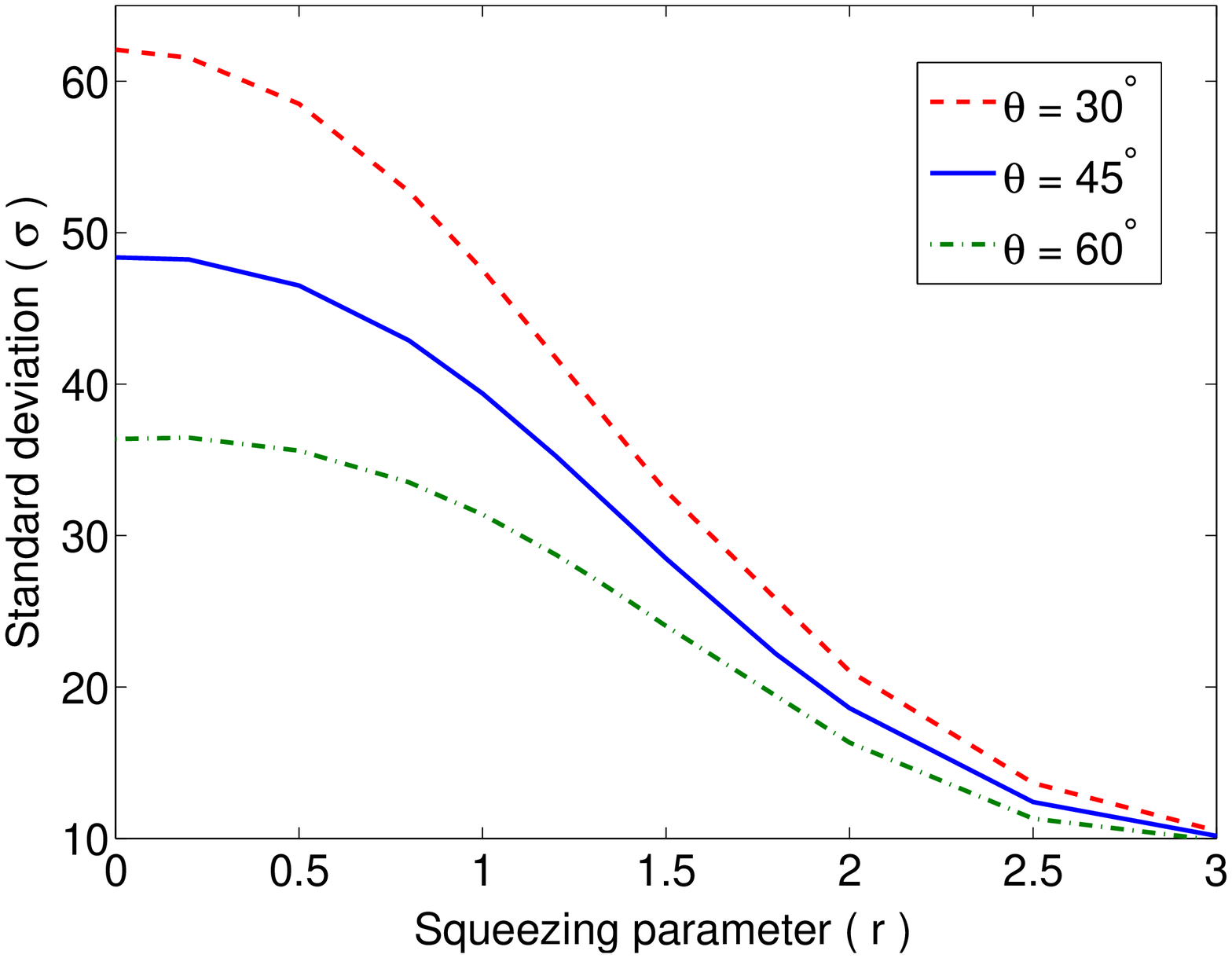}}
\caption{The effect of increasing the environmental squeezing parameter of
  the  SGAD channel,  on the  position probability  distribution  of a
  particle   with   the   initial  state   $(1/\sqrt{2})(|0\rangle   +
  i|1\rangle)\otimes |\psi_0\rangle$  and coin toss  instruction given
  by  $B(0^\circ, 45^\circ, 0^\circ)$.   Other channel  parameters are
  fixed at $T = 2$, $\gamma_{0}  = 0.025$ and $\Delta = 0.1$.  (a) The
  probability distribution on  a line for different values  of $r$, at
  time $t=100$ steps. With increasing noise (parametrized by $r$), the
  distribution  transforms  from  the  characteristic  QW  twin-peaked
  distribution to the classical  Gaussian.  (b) The Gaussianization is
  evidently accompanied  by a reduction in the  standard deviation, as
  seen here.   The plots  are given for  different values of  the coin
  toss parameter $\theta$.}
\end{figure}

\begin{figure}
\subfigure[]{\label{fig:cycsqzqw}\includegraphics[width=8.8cm]{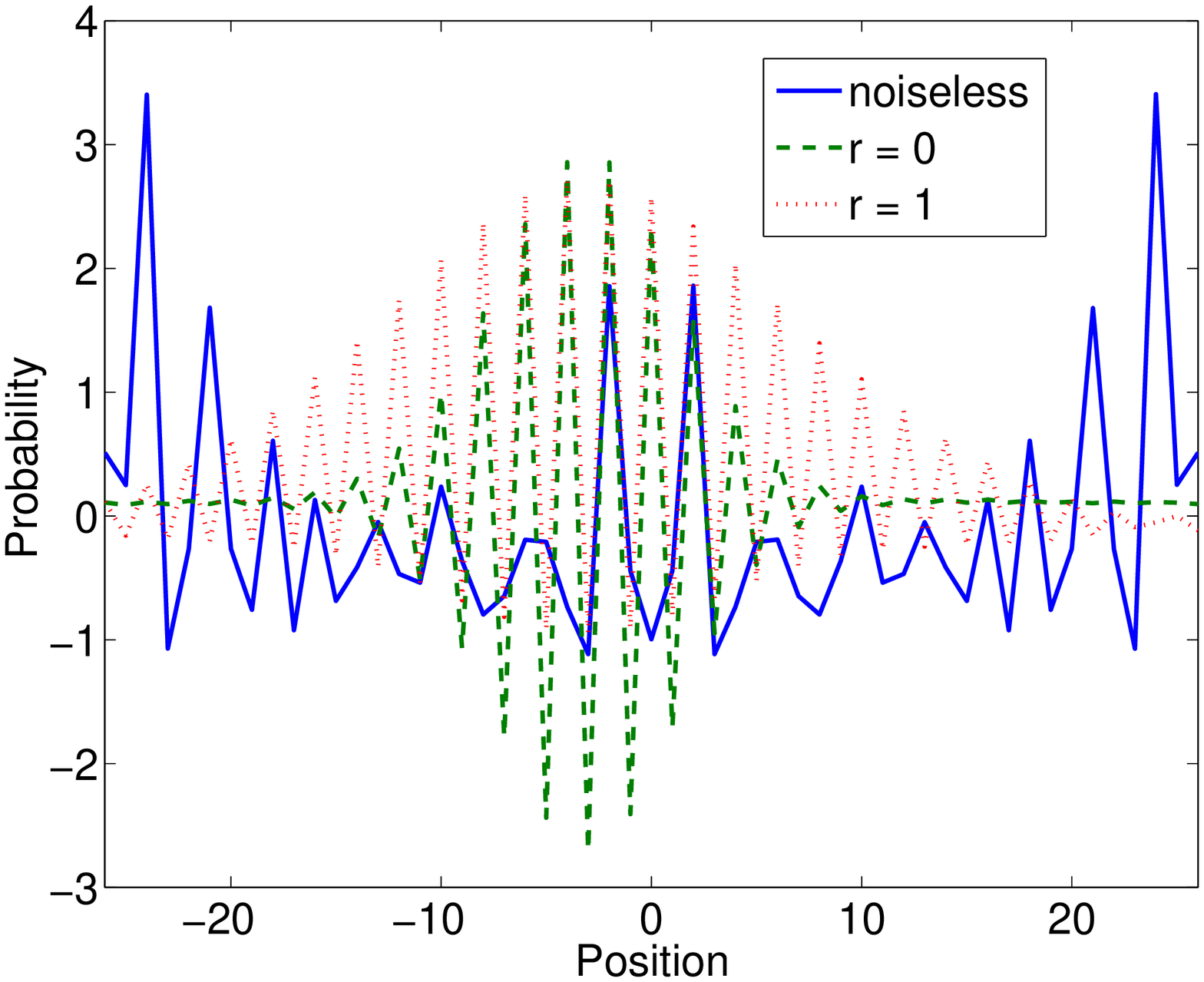}}
\hfill
\subfigure[]{\label{fig:kdcycqw}\includegraphics[width=8.8cm]{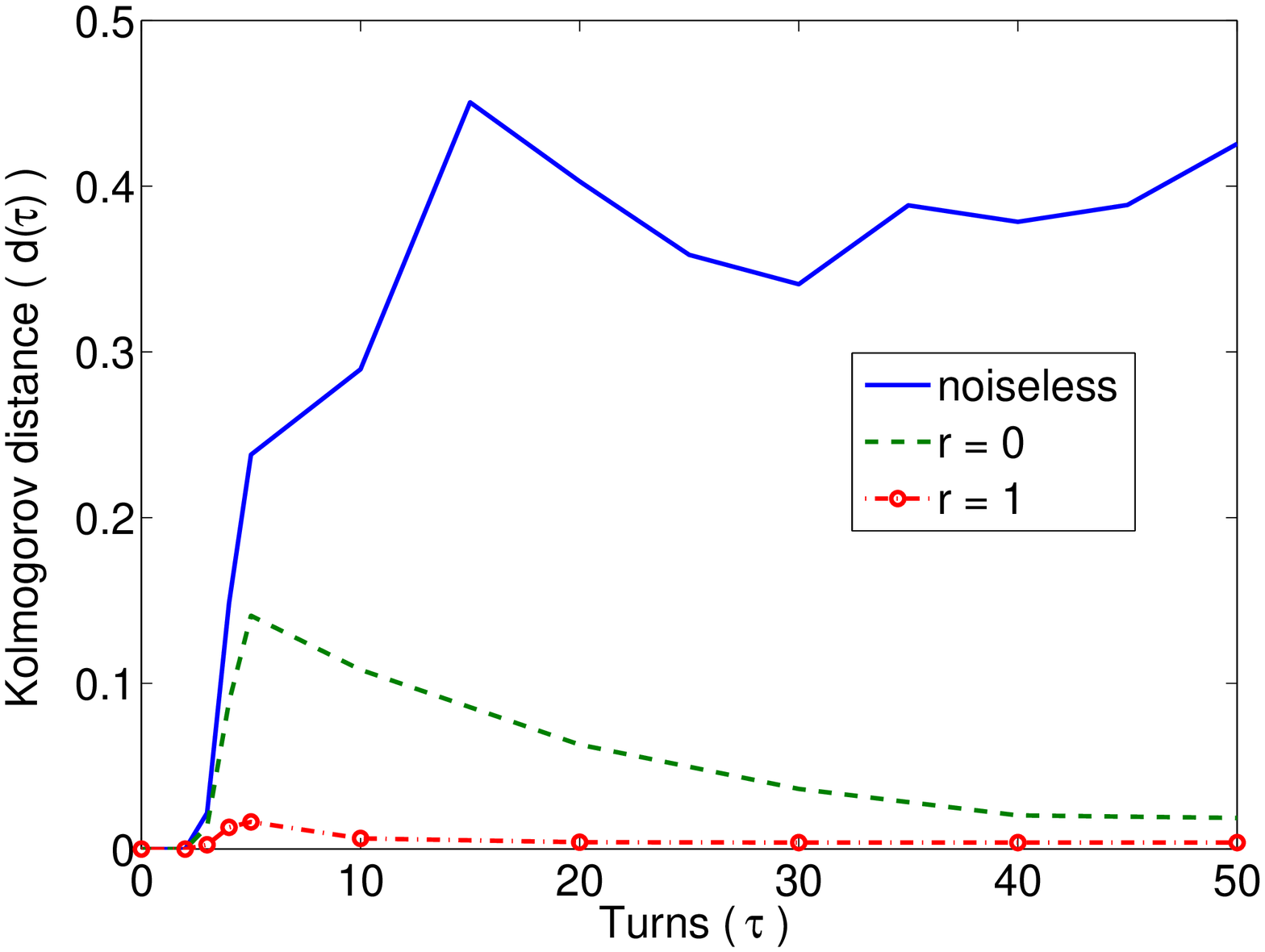}}
\caption{ (a) Probability distribution  of a particle with the initial
  state  $(1/\sqrt{2})(|0\rangle +  i|1\rangle)\otimes |\psi_0\rangle$
  and coin  toss instruction given by  $B(0^\circ, 45^\circ, 0^\circ)$
  on an $n-$cycle ($n=51$) with increase in value of the $r$ for $\tau
  =  100$.  In  contrast  to  the  case  of  QW  on  a  line  [Figs.
  \ref{fig:sqw45} and  \ref{fig:sd}], note  that  the classicalized  walk
  does  not become  Gaussian.   All  plots have  been  shifted by  the
  respective  mean value,  and downscaled  by a  factor equal  to the
  respective  standard deviation,  in  order to  bring  out the  shape
  change  in  the probability  distributions.   (b)  The breakdown  in
  `gauge  symmetry' is described  by nonvanishing  Kolmogorov (trace)
  distance  between the probability  distributions resulting  from the
  nonapplication and application of  a phase gate, $G(30^{\circ})$ at
  each  step of  the cyclic  walk \cite{BSC08}.   The unitary  walk is
  marked  by a persistent,  nonvanishing symmetry  breakdown.  Noise
  causes the gradual suppression  of superposition terms, leading to a
  restoration of the symmetry.}
\end{figure}

Figure \ref{fig:sqw45} depicts the gradual classicalization of a QW on
a  line under  the  action of  the  SGAD noise  for different  channel
parameters.   It  is   seen  that,  as  the  QW   turns  into  a  CRW,
correspondingly,  the  probability  distribution becomes  increasingly
Gaussian, causing the  quadratic functional dependence of variance 
on time, characteristic of  quantum behavior, to become linear ( cf. 
 Fig. 2 of  Ref.  \cite{BCA03}, and  the analytical  results presented  
 therein). This pattern  of quantum-to-classical  transition has also been 
 observed experimentally,  in discrete-time quantum walks using 
 single-photon, subjected to decoherence of the pure dephasing type 
  \cite{BFL10}.   This  behavior  is  quite generic, and  can be  shown, 
  for example,  to arise  under arbitrary
  Markovian decoherence  of a continuous-time quantum walk  on a graph
  (cf.  Fig. 3  of  Ref. \cite{WRA10}).   The  Gaussianization  is
directly  reflected in  the fall  of standard  deviation,  depicted in
Fig. \ref{fig:sd}, because  classicalization causes  probability to
cluster closer  to the  mean position.  As  it turns out,  neither the
Gaussianization,  nor  the   concomitant  dramatic  drop  in  standard
deviation,  happens  in  the  case  of  QW  on  an  $n$-cycle  [Fig. 
\ref{fig:cycsqzqw}].  Thus the characterization of loss of quantumness
in  a QW  in terms  of shape  change of  the  probability distribution
according to this pattern is, while convenient for walk on a line, not
fruitful in  the case  of cyclic walk.   In the following  section, we
will  discuss why a  sort of  `gauge freedom'  in QWs  also is  not an
appropriate geometry-independent measure of classicalization of QW.


%================================================
\section{A symmetry of quantum walk on a line}
\label{sec:symm}
%================================================

We will call an operation applied at each step of the walk that leaves
the  position  probability   distribution  $p(x,t)$  of  the  particle
unchanged   as   `walk   symmetry'.    The  phase   gate   $G(\alpha)=
\left(\begin{array}{ll}  1 & 0  \\ 0  & e^{i\alpha}\end{array}\right)$
acting on the coin  is such a walk symmetry for walk  on line, but not
for that on an $n-$cycle.  The physical significance of $G$ is that it
helps identify a family of  QWs that are equivalent from the viewpoint
of physical  implementation, which  can sometimes allow  a significant
practical simplification \cite{CSB07}.

Suppose   the   walker   begins    in   the   state   $(a|0\rangle   +
b|1\rangle)\otimes |\psi_0\rangle$,   with  $|a|^2+|b|^2=1$.    If   the  coin
operator is denoted  by the matrix $B \equiv  \{B_{j,k}\}$, then after
application of  this operation followed by the  conditional shift, the
walker  is  in   the  state  $\sum_{j_1}(aB_{j_10}  +  bB_{j_11})|j_1,
\psi_{2j_1-1}\rangle$, where  $j_1 \in \{0,1\}$.  Proceeding  thus, after $t$
steps  the walker  on  a line  and on  an  $n$-cycle are  seen to  be,
respectively,
\begin{subequations}
\label{eq:gsup}
\begin{eqnarray}
|\Psi_{L}(t)\rangle &=& \sum_{j_1,j_2,\cdots,j_t}e^{i J_t\alpha} 
B_{j_t,j_{t-1}}\cdots B_{j_2,j_1} (B_{j_1 0}a + B_{j_1 1}b)|j_t, \psi_{[2J_t-t]}\rangle, \label{eq:gsupa} \\
|\Psi_{C}(t)\rangle  &=&
\sum_{j_1,j_2,\cdots,j_t}e^{iJ_t\alpha}  
B_{j_t,j_{t-1}}\cdots B_{j_2,j_1} 
(B_{j_1 0}a + B_{j_1 1}b)|j_t, \psi_{[2J_t-t\mod n]}\rangle, \label{eq:gsupb}
\end{eqnarray}
\end{subequations}
where  $J_t =  j_1+j_2+\cdots+j_t$ and $0 \le J_t \le t$. The projection
of the state $|\Psi_{L}(t)\rangle$, Eq. (\ref{eq:gsupa}) onto some position state $|\psi_x\rangle \equiv |\psi_{2J_t-t}\rangle$
is given by
\begin{eqnarray}
\langle \psi_x|\Psi_L(t)\rangle &=& \sum_{j_1,j_2,\cdots,j_t}e^{i \alpha (x+t)/2} 
B_{j_t,j_{t-1}}\cdots B_{j_2,j_1} (B_{j_1 0}a + B_{j_1 1}b)|j_t\rangle \nonumber \\
&= & \sum_{j_1,j_2,\cdots,j_{[t-1]}}
\left(e^{i \alpha (x+t)/2} 
B_{0,j_{t-1}}\cdots B_{j_2,j_1} (B_{j_1 0}a + B_{j_1 1}b)|0\rangle\right. \nonumber \\
&+& \left. 
(e^{i \alpha [-1 + (x+t)/2]} 
B_{1,j_{t-1}}\cdots B_{j_2,j_1} (B_{j_1 0}a + B_{j_1 1}b)|1\rangle\right),
\nonumber 
\end{eqnarray}
where we have used the fact that $J_t = J_{t-1}+j_t = (x+t)/2$.
Therefore the probability to detect the particle at $x$ is given by:
$$
\left|\sum_{j_1,j_2,\cdots,j_{t-1}}
B_{0,j_{t-1}}\cdots B_{j_2,j_1} (B_{j_1 0}a + B_{j_1 1}b)\right|^2 +
\left|\sum_{j_1,j_2,\cdots,j_{t-1}}
B_{1,j_{t-1}}\cdots B_{j_2,j_1} (B_{j_1 0}a + B_{j_1 1}b)\right|^2,
$$ independent of $\alpha$, whence the symmetry.  Proceeding similarly
in  the  case  of QW  on  an  $n$-cycle,  we  find that  whereas  each
superposition  term accumulates phase  in the  same way,  the spatial
spread of the  state is restricted by the geometry  not to exceed $n$,
leading to Eq.  (\ref{eq:gsupb}).  As  a result, fixing $x$ fixes $J_t
\mod n  = (x+t)/2  \mod n$, but  not $J_t$  itself, so that  the phase
terms  in the superposition  Eq.  (\ref{eq:gsupb})  do not  factor out
globally  in  the above  manner  as  with QW  on  a  line, whence  the
breakdown    in   the    symmetry   in    the   cyclic    case   [Fig. 
\ref{fig:kdcycqw}].  We use the  notation $n=2s+1$,  where $s$  is the
number of vertices  on the left or right arm  of the cycle, excluding
the origin.  We define turns $\tau$  by $t=\tau s$, where  $t$ is walk
time or  the number of  steps. Physically, the breakdown  in symmetry
can be attributed to  interference of the clockwise and anticlockwise
rotating waves on the cycle.

In Eq.  (\ref{eq:gsupa}), note that  the proof of symmetry  works just
the same even  if the $B$'s in each  operation are distinct operators,
and,  further, even if  they are  arbitrary matrices,  not necessarily
unitary. With this insight, consider any CP map represented by the set
of Kraus  operators $\{E_1,\cdots,E_d\}$.  The operators $E_j$  may be
unitary, as with a Pauli channel like bit flip or phase flip channels,
or not, as with the  amplitude damping channel or squeezed generalized
amplitude damping  channel.  Consider  an $n$-fold application  of the
noise  process represented by  this CP  map on  an initial  pure state
$\rho \equiv |\Psi_{in}\rangle\langle\Psi_{in}|$. This produces
\begin{equation}
\rho^\prime = \sum_{i_1,\cdots,i_n=1}^d E_{i_n}\cdots E_{i_1} \rho
E^\dag_{i_1}\cdots E^\dag_{i_n}
\end{equation}
The  state $\rho^\prime$  can equally  well be  regarded as  a uniform
mixture of  the $d^n$ unnormalized  states of the  form $E_{i_n}\cdots
E_{i_1}|\Psi_{in}\rangle$,  where  a  particular sequence  ${\bf  E}_k
\equiv   E_{i_n}\cdots  E_{i_1}$   can   be  regarded   as  a   `noise
trajectory'.  Note  that  the  $\{{\bf E}_k\}$  are  themselves  Kraus
operators for they can be shown to be positive and to satisfy
\begin{equation}
\sum_{k=1}^{d^n} {\bf E}^\dag_k {\bf E}_k = \mathbb{I}. \nonumber \\
\end{equation}

A  particular  noise trajectory  of  $n$ steps  of  a  walk, given  by
$(UB)^n|\Psi_{in}\rangle$       could      be,       for      example,
$(UBE_{i_n})(UBE_{i_{n-1}})\cdots(UBE_{i_2})(UBE_{i_1})|\Psi_{in}\rangle$,
where  the  noise is  assumed  to  act on  the  coin  just before  the
application   of  $B$   at  each   step.   This   can   be  rewritten,
matrix-multiplying   $B$   with   the   noise   operator,   as   ${\bf
  E}_x|\Psi_{in}\rangle                                          \equiv
(UB^\prime_{i_n})(UB^\prime_{i_{n-1}})\cdots(UB^\prime_{i_2})(UB^\prime_{i_1})|\Psi_{in}\rangle$.
If the $B$'s in Eq. (\ref{eq:gsupa}) are replaced by the $B^\prime$'s,
we  are  led  by  the  same  reasoning to  conclude  that  along  each
trajectory,   $p(x,t)$   is   unaffected   by   the   application   of
$G(\alpha)$. Hence, the average  overall $d^n$ noise trajectories will
also leave $p(x,t)$ unaffected.

We assumed above that the noise acts at each step just before the coin
operation, but it is clear that it may equally well act just after the
coin  operation.  Furthermore,  it   may  also  act  after  the  shift
operation. To see  this we rewrite, using the  associative property of
quantum   operations,   a   particular  noise   trajectory   $(E_{i_n}
UB)(E_{i_{n-1}}UB)\cdots(E_{i_2}UB)(E_{i_1}UB)|\Psi_{L}\rangle$      as
$E_{i_n}                                       (UBE_{i_{n-1}})(UB\cdots
E_{i_2})(UBE_{i_1})UB|\Psi_{in}\rangle$,      then     as     $E_{i_n}
(UB^\prime_{i_{n-1}})(\cdots
UB^\prime_{i_2})(UB^\prime_{i_1})UB|\Psi_{L}\rangle$  and  finally  as
$E_{i_n}                   (UBE_{i_{n-1}})(UB                   \cdots
E_{i_2})(UBE_{i_1})|\Psi_{L}^\prime\rangle$  and once again  the above
reasoning can  be applied to show  that symmetry holds  even with this
model of noise.  In earlier
works, we  explicitly verified this for  the case when the  noise is a
phase damping channel \cite{CSB07} and a generalized amplitude damping
channel \cite{BSC08}. Not  surprisingly, noise {\it restores} symmetry
for the  cyclic QW, which can  be understood simply  as the
result of suppression of quantum superpositions \cite{BSC08}.

Figure  \ref{fig:kdcycqw} depicts  the breakdown  of the  symmetry in
noiseless QW on an $n$-cycle, and its restoration with the application
of noise. At each time  step, symmetry breakdown is quantified as the
trace or  Kolmogorov distance \cite{NC00} between  the two probability
distributions obtained by  applying or not applying the  phase gate on
every step. We  note that while $\tau \le 2$,  the cyclic walk remains
symmetric, as  the diffusing particle  is `unaware' of  its nontrivial
topology.  From  Eq.  (\ref{eq:gsupb}), we  see that the  beginning of
breakdown is when $\max(2J_t-t) \ge  n$, which happens when $t\ge n$,
or $\tau \ge 2 + (1/s)$.  Clearly, even though the symmetry breakdown
is  potentially a  good probe  of walk  topology, as  an  indicator of
quantumness,  it is  relevant  only to  cyclic  walk, and  thus not  a
geometry-independent indicator of classicalization.

%========================================
\section{Quantumness using measurement-induced disturbance}
\label{MID}
%========================================

Finally  we consider  the  classicalization of  QW  on a  line and  an
$n-$cycle under  the influence  of the SGAD  channel, using QMID  as a
measure  of  quantumness of  the  correlations  between  the coin  and
position degrees of freedom.
\begin{figure}
\subfigure[]{\label{fig:qness1}\includegraphics[width=8.8cm]{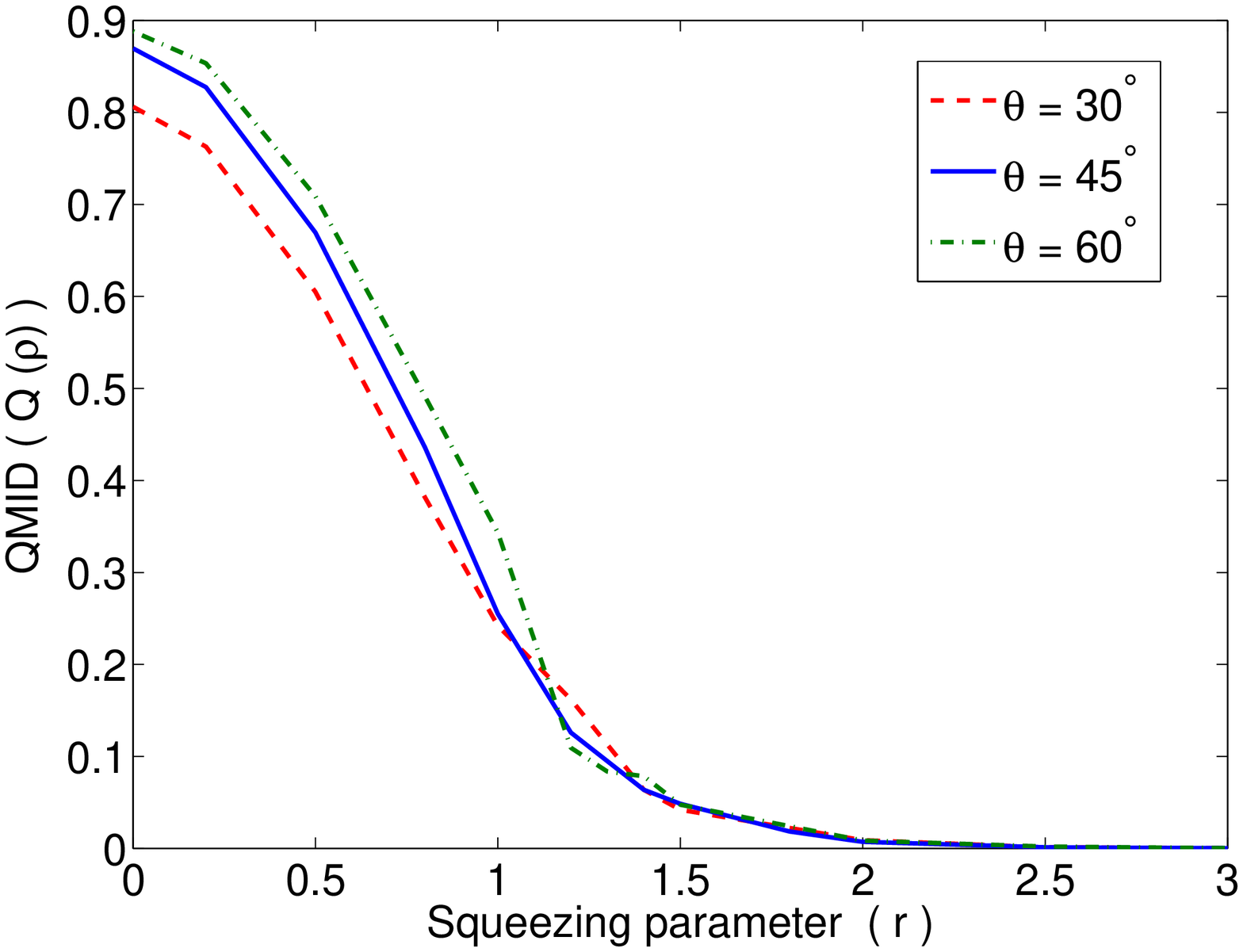}}
\hfill
\subfigure[]{\label{fig:qness2}\includegraphics[width=8.8cm]{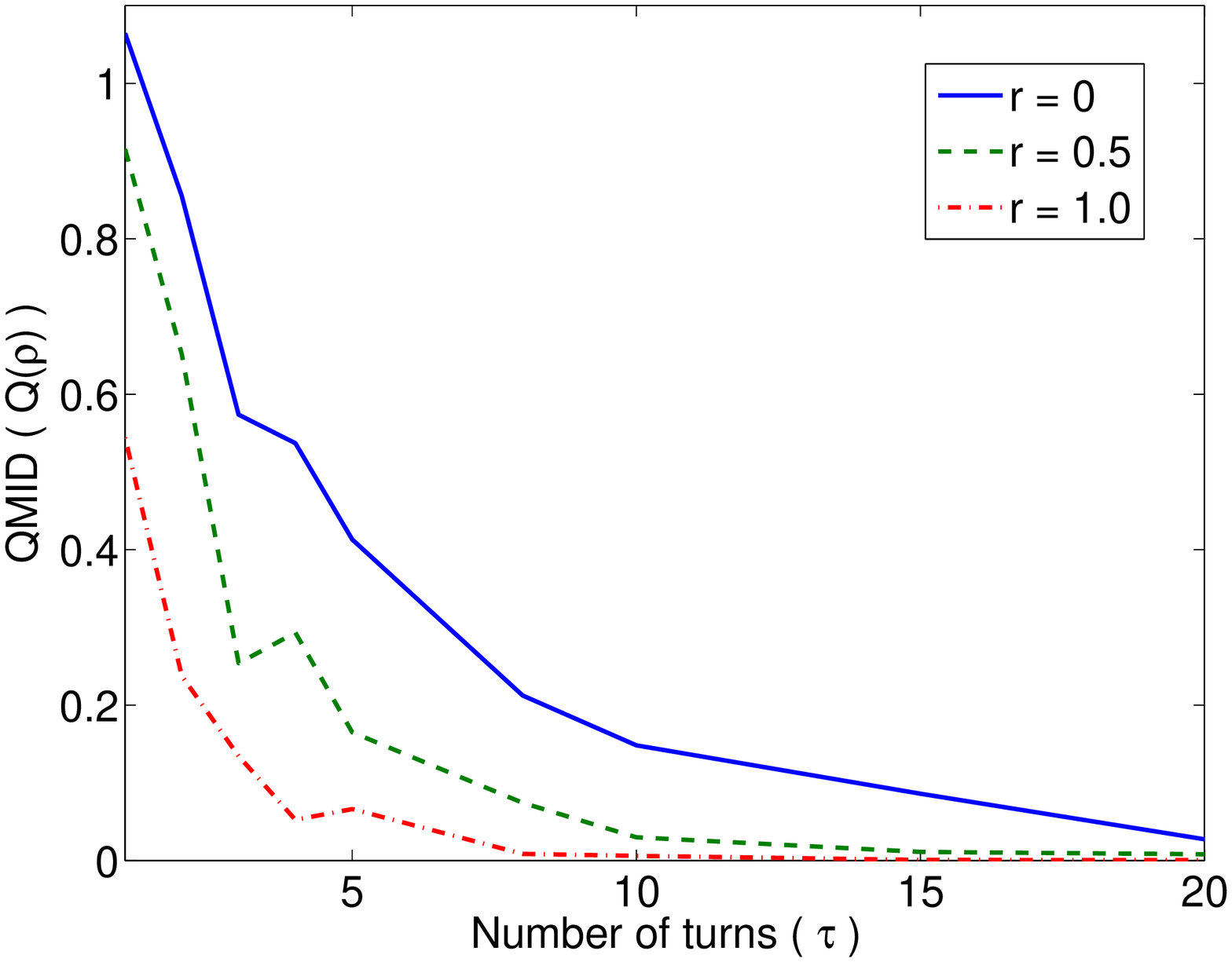}}
\caption{ (a) Decrease in QMID with increasing squeezing parameter $r$
  for a  particle walking on  a line at  time $t=100$ steps,  with the
  initial state  $(1/\sqrt{2})(|0\rangle + i|1\rangle)\otimes |\psi_0\rangle$  and coin toss
  instruction given  by $B(0^\circ, \theta,  0^\circ)$, where $\theta=
  30^\circ$,  $\theta=  45^\circ$, and  $\theta=  60^\circ$. A  similar
  decreasing pattern  can also be seen  in the cyclic  case.  Note the
  slight $\theta$ dependence of $Q$, and the switch in this dependence
  around $r=1.2$.  (b) Variation of QMID $[Q(\rho)]$ for QW on an
  $n-$ cycle ($n=101$), with increase in number of turns ( $\tau$ ) of
  the  cycle.   With  increase  in $r$,  $Q(\rho)$  decreases  faster,
  implying   more   rapid   classicalization.    Other   environmental
  parameters  are  $\gamma_{0}  =  0.025$, temperature  ($T  =2$),  and
  $\Delta = 0.1$.}
\label{fig:sdqness1}
\end{figure}
Given a bipartite state $\rho$ living in the Hilbert space ${\cal H}_A
\otimes {\cal H}_B$, and the reduced density matrices being denoted by
$\rho_A$  and $\rho_B$,  a  reasonable measure  of total  correlations
between systems $A$ and $B$ is the mutual information:
\begin{equation}
I(\rho) = S(\rho_A) + S(\rho_B) - S(\rho),
\label{eq:I}
\end{equation}
where  $S(\cdot)$ denotes  von Neumann  entropy. If  $\rho_A  = \sum_j
p_A^i\Pi_A^i$ and $\rho_B = \sum_j p_B^i\Pi_B^i$, then the measurement
induced by the spectral resolution of the reduced states is
\begin{equation}
\label{eq:Pi}
\Pi(\rho) \equiv \sum_{j,k} \Pi_A^j \otimes \Pi_B^k \rho
\Pi_A^j \otimes \Pi_B^k.
\end{equation}
The state $\Pi(\rho)$ may be  considered classical in the sense that there
is a  (unique) local measurement  strategy, namely $\Pi$,  that leaves
$\Pi(\rho)$ unchanged. This strategy is  special in that it produces a
classical state in $\rho$ while keeping the reduced states invariant.

According to Luo  \cite{L08}, we may consider as  a measure of quantum
correlations the quantity $D[\rho,\Pi(\rho)]$, where $D$ is a suitable
measure of quantum distance, such as Hilbert-Schmidt distance or Bures
measure. If  we accept that $I[\Pi(\rho)]$  is a good  measure of {\it
  classical} correlations in $\rho$, then one may consider QMID, given
by
\begin{equation}
\label{eq:Q}
Q(\rho) = I(\rho) - I[\Pi(\rho)],
\end{equation}
as a reasonable measure of quantum correlation \cite{L08}.

Our  application  of   QMID  to  QW  to  derive   the  quantumness  of
coin-position  correlations   supports  this  view.    We  numerically
determined $\rho$ in Eq. (\ref{eq:I}),  by applying SGAD noise at each
walk  step,  for  various  channel parameters  like  temperature  $T$,
coupling  constant $\gamma_0$,  $\Delta$, and  environmental squeezing
parameter $r$.   The operator $\Pi$  in Eq. (\ref{eq:Pi}),  and thence
$Q(\rho)$ in Eq. (\ref{eq:Q}),  are obtained by diagonalization of the
reduced density matrices $\rho_A$ and $\rho_B$.

\begin{figure}
\subfigure[]{\label{fig:lineVSncycle1}\includegraphics[width=8.8cm]{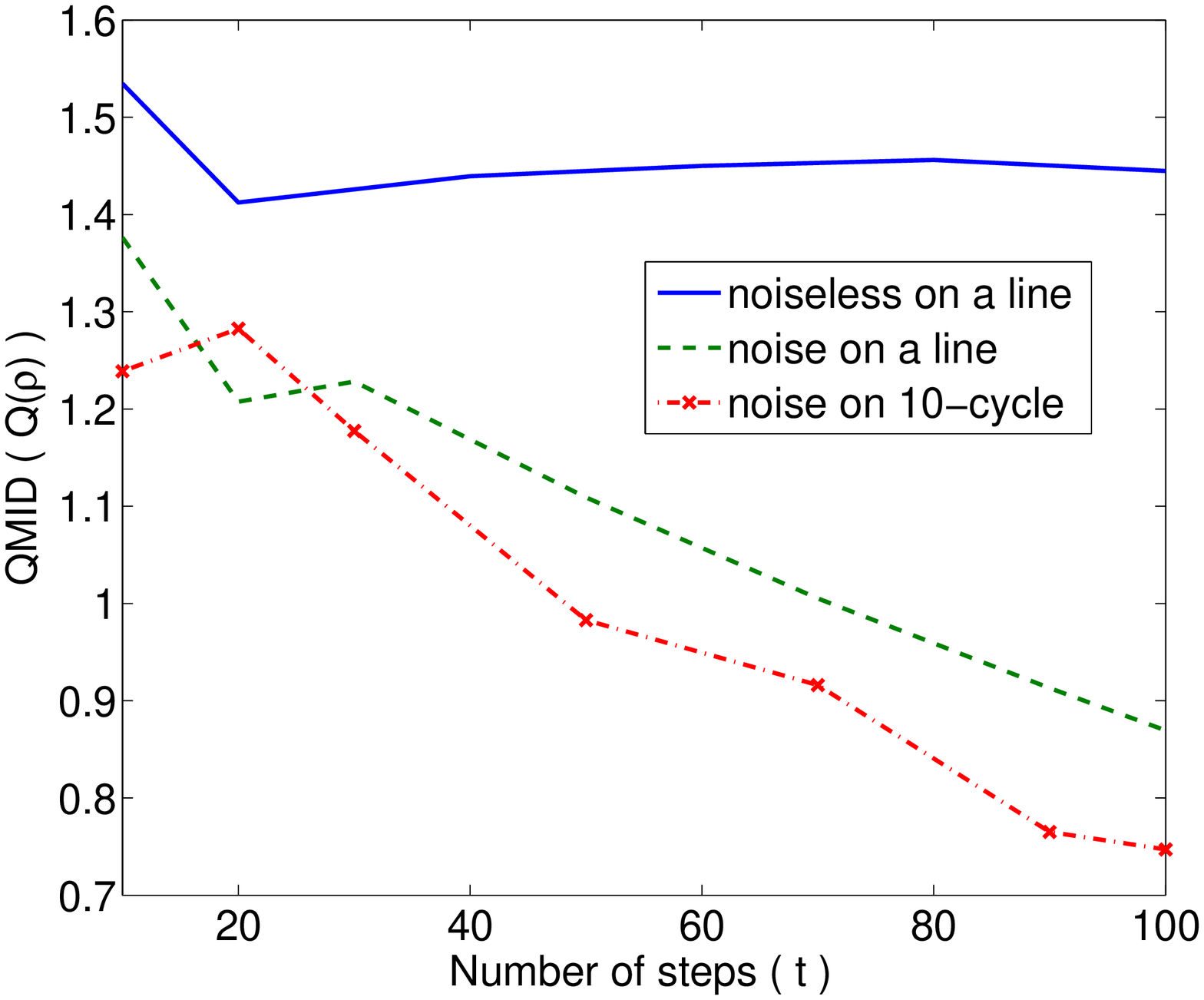}}
\hfill
\subfigure[]{\label{fig:lineVSncycle2}\includegraphics[width=8.8cm]{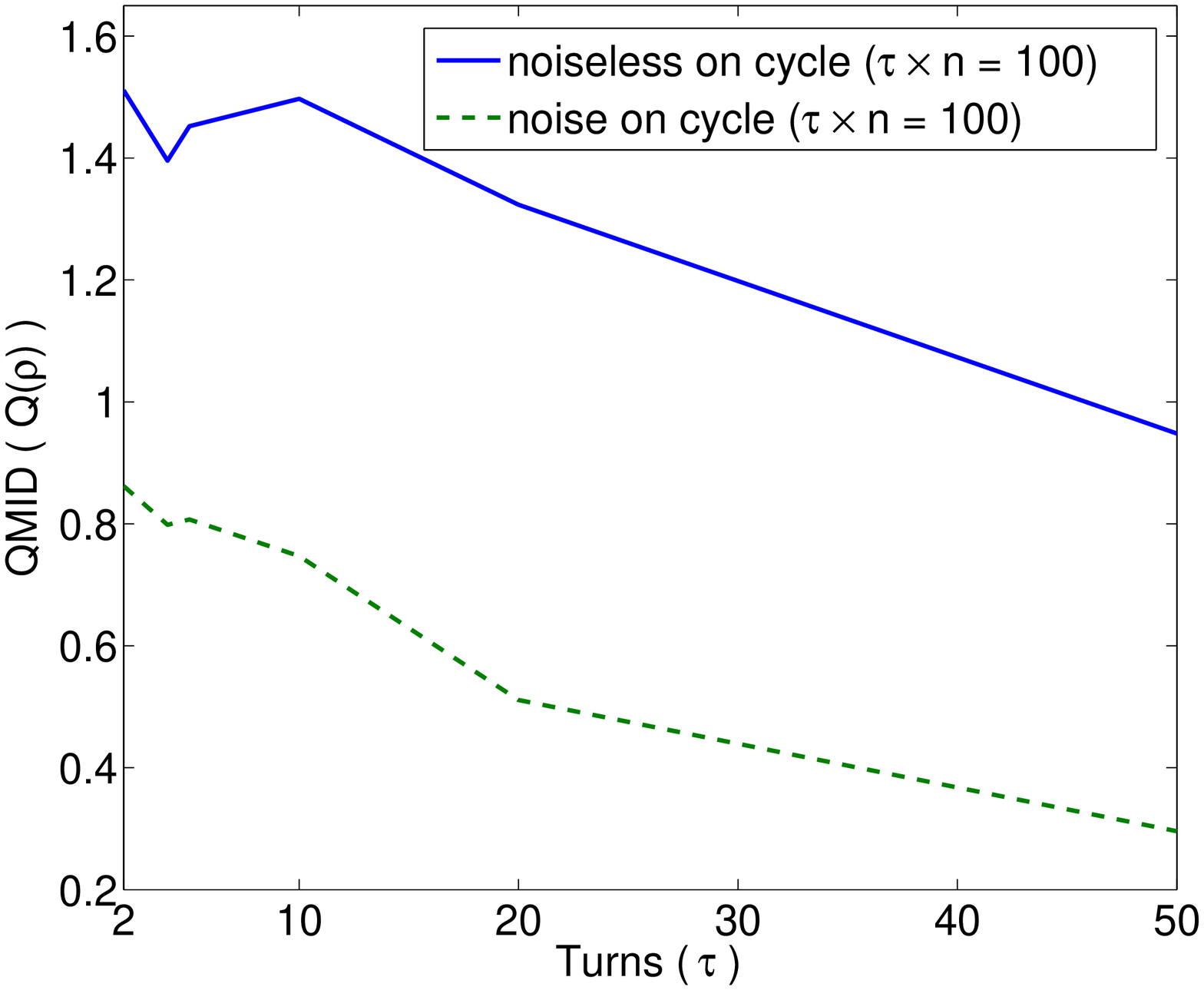}}
\caption{A comparison of the  degree of classicalization between QW on
  a  line  and that  on  an  $n$-cycle, via  QMID  as  the measure  of
  quantumness.   (a) For  a fixed  setting of  the environment-induced
  decoherence (here $T=0.1$, $\gamma_0=0.05$, $\Delta=0.1$  and $r=0$) and any given
  sufficiently large time, the cyclic walk typically becomes classical
  more  rapidly  on  account  of  enhanced randomization  due  to  the
  interference of  the counter-rotating waves.  (b)  This is evidenced
  by  the suppression of  QMID with  increase in  the number  of turns
  $\tau$, keeping time $t$ and all noise parameters fixed.}
\label{fig:lineVSncycle}
\end{figure}

Figure \ref{fig:sdqness1} presents typical plots  of QMID in a QW system
subjected  to a SGAD  channel  with different  variable parameters.   In
Fig.  \ref{fig:qness1}, we  depict the fall of QMID in  noisy QW on a
line  with increasing  environment  squeezing $r$.  A similar  pattern
emerges  also  in the  case  of  cyclic  QW.  Figure  \ref{fig:qness2}
depicts the fall of QMID in a noisy, cyclic QW with increasing time, a
pattern that is  seen also in the  case of QW on a  line. Both figures
demonstrate,  as   expected,  that  increasing  the   level  of  noise
(parametrized   by   $r$)   is    detrimental   to   the   degree   of
quantumness. More generally, these findings suggest that QMID provides
the desired  geometry-independent way to  describe the classicalization
of QW, unlike the previously discussed ways.

With this result, we may compare  the decohering properties of QW on a
line with those of  cyclic QW (Fig. \ref{fig:lineVSncycle}). For fixed
noise  setting, at  any given  sufficiently large  time, as  seen from
Fig. \ref{fig:lineVSncycle1}, cyclic walk  is more classical than walk
on  the  line,  which  may  be  understood  as  follows.   Decoherence
randomizes phase, and in cyclic walk, the self-interference of the two
oppositely   rotating  waves   enhances   this  randomization.    This
explanation is  supported by Fig. \ref{fig:lineVSncycle2}, where
increasing the number  of turns $\tau$ keeping time  $t$ and all noise
parameters fixed is seen to suppress the quantumness.

%===================================
\section{Conclusions} \label{conclusion}
%===================================

In this work, we investigated  the loss of quantumness in a decoherent
DTQW  on a  line and  an  $n$-cycle. We  showed that  QMID provides  a
reasonable   geometry-independent  description   of   the  degree   of
classicalization of  the walk, unlike  such familiar criterion  as the
Gaussianization  of the walk  probability distribution,  which applies
only  to walk  on  a line,  or  a specific  symmetry principle,  which
applies  only to  cyclic  walks.  The  noise  is modeled  as an  SGAD
channel  acting  on  the  coin.   One might  consider  the  amount  of
entanglement between the coin and the position degrees of freedom as a
measure of  quantumness, but, as  mentioned in the  Sec. \ref{sec:intro}, this
might be an underestimate  because it ignores the non-classicality due
to non-commutative observable algebra of the coin and position degrees
of freedom.  Quantum discord  is a transparent measure of quantumness,
but its  computation for  large systems like  a quantum walker  can be
hard because of the optimization it involves.  This suggests that QMID
is a  suitable, geometry-independent operational tool  to describe the
quantumness of QW. As a first application of this idea, we showed that
cyclic QWs tend  to classicalize faster than QWs on  a line on account
of more efficient phase randomization.

%=================================

\end{document}